\documentclass[a4paper,11pt]{article}
\pdfoutput=1 

\usepackage{jcappub} 

\usepackage[T1]{fontenc} 

\usepackage{multirow} 
\usepackage{amsmath}
\usepackage{times}

\title{\boldmath 
The scale of homogeneity in the local Universe with the ALFALFA catalogue
}

\author[a]{Felipe Avila,}
\author[a]{Camila P. Novaes,}
\author[a]{Armando Bernui,}
\author[a,b]{Edilson de Carvalho}

\affiliation[a]{Observat\'orio Nacional, \\
Rua General Jos\'e Cristino 77, S\~ao Crist\'ov\~ao, 20921-400, Rio de Janeiro, RJ, Brazil}
\affiliation[b]{Centro de Estudos Superiores de Tabatinga, Universidade do Estado do Amazonas, \\
69640-000, Tabatinga, AM, Brazil}

\emailAdd{felipeavila@on.br}
\emailAdd{camilapnovaes@gmail.com}
\emailAdd{bernui@on.br}
\emailAdd{edilsonfilho@on.br}

\abstract{We use the scaled counts in spherical caps $\mathcal{N}(<\theta)$ and the fractal correlation dimension $ \mathcal{D}_{2}(\theta) $ procedures to search for a transition scale to homogeneity in the local universe as given by the ALFALFA catalogue (a sample of extragalactic HI line sources, in the redshift range $0 < z < 0.06$). 
Our analyses, in the 2-dimensional sky projected data, show a transition to homogeneity at $\theta_H = 16.49^{\circ} \pm 0.29^{\circ}$, in remarkable accordance with the angular scale expected from simulations, a result that strengthens the validity of the cosmological principle in the local universe. 
We test the robustness of our results by analysing the data sample using three versions of the $\mathcal{N}(<\theta)$ estimator, which show a well agreement between them. 
These statistical estimators were validated using mock realizations generated assuming a fractal distribution of points, successfully recovering the input information. 
In addition, we perform further analyses showing that our approach is also able to indicate the presence of under- and over-densities in the data.
Finally, we verify the influence of the sample size to the $\theta_H$ estimates by using segment Cox processes of different projected areas, confirming the suitability of the ALFALFA surveyed area for the current analyses.
}

\begin{document}
\maketitle
\flushbottom

\section{Introduction} \label{sec1}
\paragraph{}
The cosmological principle (CP) is at the basis of modern cosmology~\cite{peacock,Maartens11}. 
The CP, together with Einstein's equations, supports the concordance cosmological model, 
$\Lambda$CDM, which currently shows the best agreement with observational data (see, 
e.g.,~\cite{review1,review2} for a review of the concordance model and its key results). 
According to the CP, the Universe is expected to be homogeneous and isotropic at sufficiently 
large scales. 
With the advent of new and precise astronomical data, it becomes an open problem 
to estimate how large are these scales~\cite{Clarkson12,Heavens11,Valkenburg}.

Nowadays, the study of the statistical isotropy of matter and radiation is an active research field. 
In the absence of information to know the cosmological distances, the {\em projected} 
isotropy~\cite{Laurent}, i.e., considering the sources projected on the sky, has been extensively tested 
with extra-galactic sources like X-ray~\cite{raiox}, Radio~\cite{radio,radio2}, Gamma-ray 
bursts~\cite{BFW,isogamma,Tarnopolski,Ripa}, galaxy clusters~\cite{BengalyPSZ} and full sky galaxy 
survey~\cite{Pandey2017,Carneiro}, showing their compatibility with statistical isotropy. 
Besides, the {\em projected} isotropy of the Planck convergence map~\cite{GAM} and of the WISE galaxy map~\cite{BengalyWISE,Yoon,Bengaly18,Novaes18} have been also investigated without indications of 
a possible CP violation. 
The cosmic microwave background (CMB) radiation is also consistent with statistical isotropy at small 
angular scales~\cite{PLA1-XXIII,PLA2-XVI,CPN}, although some controversy exists at the largest 
scales~\cite{Gruppuso,Polastri,Schwarz,BOP,Aluri,Rath,BNPS,AB,Abramo}.

The analysis of spatial homogeneity is more cumbersome. 
Methods that test the homogeneity of the matter distribution by counting objects in spheres or 
spherical caps (when the sources are projected on the sky) are not direct tests of spatial homogeneity. 
The information regarding the number of cosmic objects on spatial hyper-surfaces inside the past 
light-cone cannot be accessed by methods like {\em counts-in-spheres} because the counts are restricted 
to the intersection of the past light-cone with the spatial hyper-surfaces (see, e.g.,~\cite{Maartens11,%
Clarkson12}). 
However, these methods provide consistency tests in the following sense: if the counts-in-spheres method 
shows that the objects distribution does not approach homogeneity on large scales, then this 
can falsify the CP. 
Instead, if observations confirm the existence of a transition scale to homogeneity, then this strengthens 
the evidence for spatial homogeneity, but cannot prove it~\cite{Maartens11,Heavens11}.

The main difficulty to perform consistency tests concerns the low number density of cosmic objects 
achieved in past surveys, insufficient to obtain a good signal to noise ratio. 
Recent surveys, however, increased substantially the volume of the observed universe and the 
purity of the cosmic tracers samples, promoting this type of analysis.

The counts-in-spheres method is being widely used to explore the homogeneity scale in 
diverse cosmological tracers. 
It is based on the idea that, for a homogeneous sample, when averaging the normalized number 
of cosmic objects inside spheres of radius $r$, $N(< r)$, one should observe a behaviour 
like $N(<r) \propto r^3$ when increasing the sphere radius $r$ as large as possible. 
This estimator gives a relatively good measure for real surveys with peculiar geometry, but its 
correlation matrix reveals that their measurements are highly correlated. 
Alternatively, one can use the {\em fractal correlation dimension}, $\mathcal{D}_{2}(r)$, 
which has the advantage to be essentially uncorrelated between bins of 
distance~\cite{Laurent,Scrimgeour,Ntelis}, besides to correct some systematic effects like 
holes or gaps in the survey geometry, the catalogue incompleteness, and needs not to assume 
homogeneity at large scales as required in the correlation function estimator~\cite{Coleman}.

Applying the scaled counts-in-spheres and the fractal correlation dimension procedures in 
3-dimensional (3D) analyses, it was recently reported the transition scale to homogeneity 
for blue galaxies from the WiggleZ survey, with redshifts $0.1 < z < 0.9$~\cite{Scrimgeour}. 
More recently it was also applied to galaxy catalogues~\cite{Sarkar,Ntelis} and to quasars 
sample~\cite{Laurent,Nada,rodrigoquasar,Ntelis2} of the Sloan Digital Sky Survey (SDSS). 
In addition, it was used to confirm that the transition scale varies as a function of the 
universe age~\cite{Ntelis,Rodrigo}: the higher the redshift the less clumpy is the matter. 
However, despite the robustness of these studies, they used the standard cosmological 
model to calculate distances which, in some way, may bias the results. 
With this in mind, some authors~\cite{Alonso14,Rodrigo,Alonso15} worked on an independent model 
analysis, studying the homogeneity with projected 2D data.

In this work we search for the angular scale of transition to homogeneity in the 2D sky projected data 
using the ALFALFA catalogue. 
The ALFALFA project is a radio survey that observed extragalactic HI line sources of the local universe, 
$0 < z < 0.06$. 
We use these extraordinary data to explore the homogeneity scale in the local universe because the 
recently released catalogue is 100\% complete and has a very good ratio between surveyed area and 
detected number of objects~\cite{haynes2018,giovanelli} (for other applications of this dataset see, 
e.g.,~\cite{zu,jones}).

This work is organized as follows. 
In section 2 we present briefly some features of the survey. In section 3 we describe the method of 
scaled counts-in-spheres and its definitions. 
In sections 4 and 5 we show our results and present our conclusions, respectively.

\section{The Arecibo Legacy Fast ALFA Survey} \label{sec2}
\paragraph{}

The Arecibo Legacy Fast ALFA Survey\footnote{\url{http://egg.astro.cornell.edu/alfalfa/data/index.php}} (ALFALFA) is a catalogue of extragalactic HI line sources, in 21 cm, which cover an area of $ \sim 7000 \,\text{deg}^{2}$. 
Most of the sources detected are gas-rich galaxies with low surface brightness and dwarfs that populate the local universe $0 < z < 0.06$~\cite{haynes,haynes2018}. 
Recently, it has been released the first version of the full ALFALFA catalogue\footnote{The catalogue analysed here was kindly provided by M. P. Haynes, on behalf of the ALFALFA collaboration, before the public release.}, with 100\% of the footprint area, containing $33573$ objects classified in three categories according to its HI line detection status: 
{\sc code} 1, high signal to noise ratio extragalactic sources, considered highly reliable and with confirmed optical counterpart;
{\sc code} 2, lower signal to noise ratio HI signal coincident with optical counterpart, considered unreliable sources; and 
{\sc code} 9, high signal to noise ratio source with no optical counterpart and likely Galactic high velocity cloud. 
In this work we shall use only the sources designated as {\sc Code 1}, as recommended by the ALFALFA team. 
The ALFALFA survey covers two continuous regions, both in the declination range 
$0^{\circ} < \text{DEC} < 36^{\circ}$, in the right ascensions intervals of 
$21^{\text{h}} 30^{\text{m}} < \text{RA} < 3^{\text{h}} 15^{\text{m}}$ and 
$7^{\text{h}} 20^{\text{m}} < \text{RA} < 16^{\text{h}}40^{\text{m}}$. 
The first region, referred as {\it Fall sky}, contains a total of 11433 HI sources and the second one, the 
{\it Spring sky}, have 22140 HI sources. 

\begin{figure*}[!h]
\begin{center}
\mbox{\hspace{-0.3cm}
\includegraphics[width=8cm, height=5.8cm]{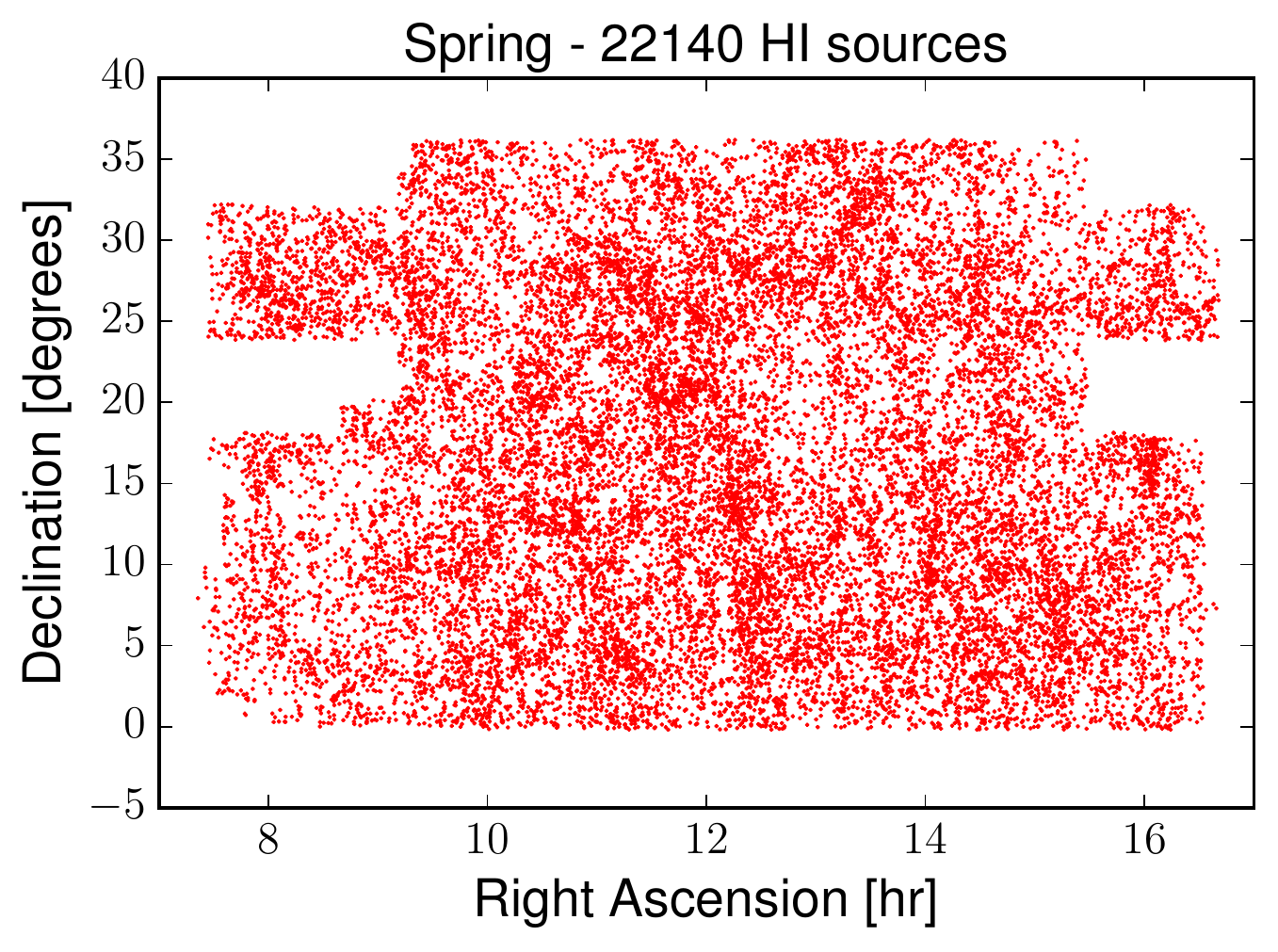}
\hspace{-0.3cm}
\includegraphics[width=8cm, height=5.8cm]{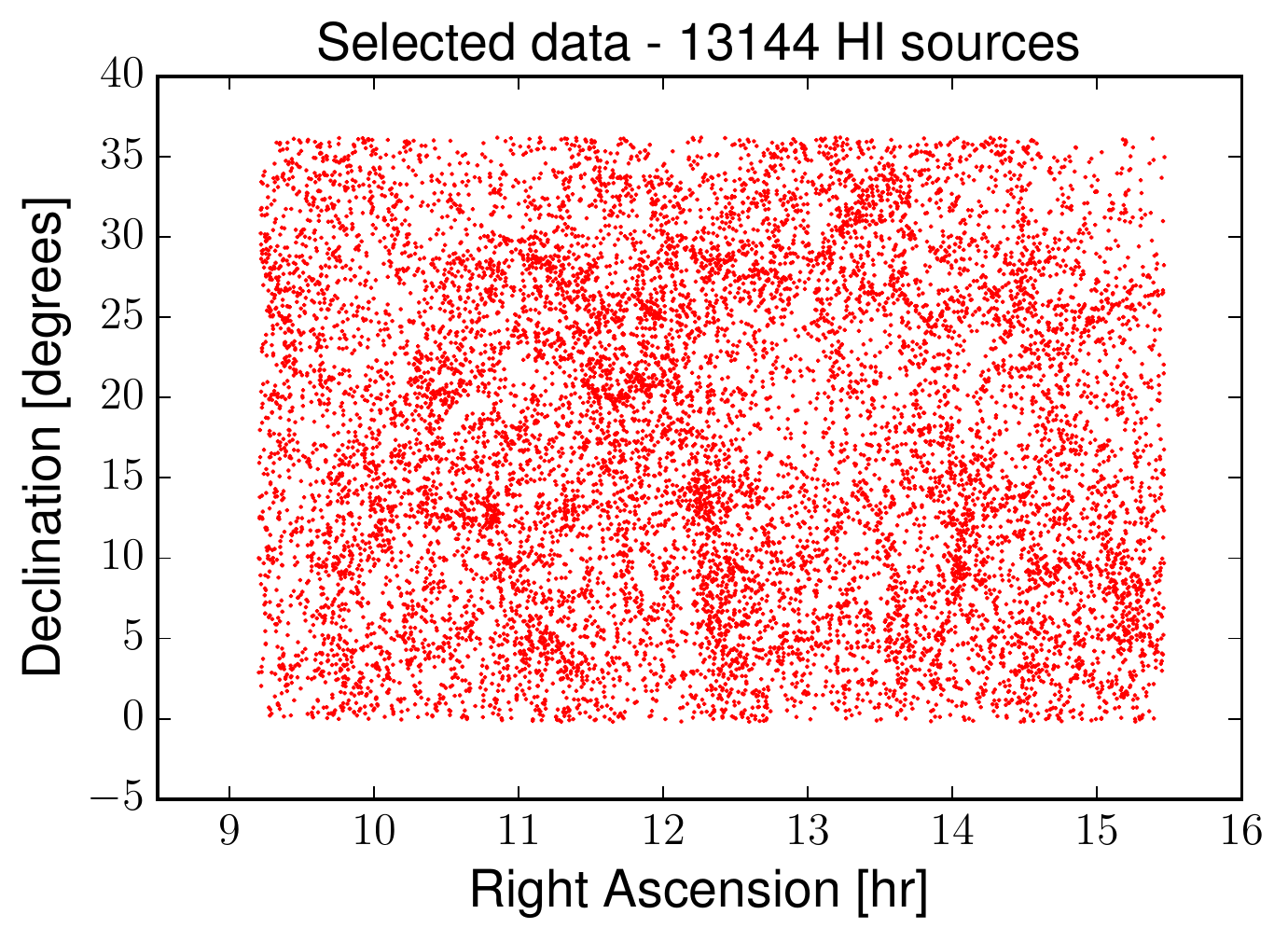}
}
\caption{{\bf Left:} Cartesian projection of the area corresponding to the Spring sky region. 
{\bf Right:} Cartesian projection of the final selected sample for analyses. 
Note the cuts in RA on the left and right sides of the Spring sky to minimize possible systematics 
introduced by the irregular geometry of the survey.
}
\label{fig1}
\end{center}
\end{figure*}

\begin{figure}[!h]
\centering
\includegraphics[width=8cm, height=5.8cm]{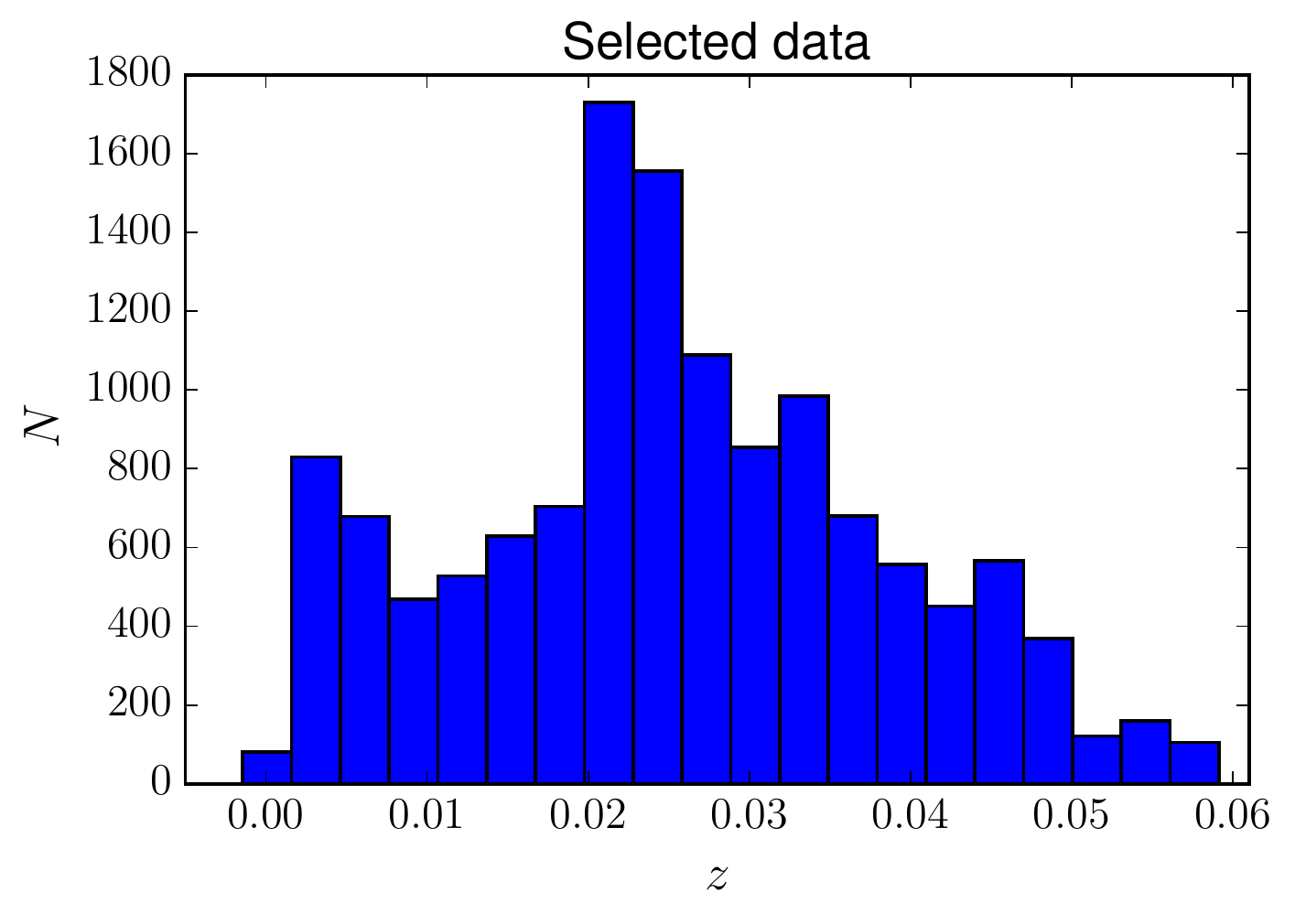}
\caption{Redshift distribution of sources in the final sample selected from the ALFALFA catalogue (corresponding to the right panel of figure \ref{fig1}). 
}
\label{fig2}
\end{figure}

Here we consider only the Spring sky region, illustrated in the left panel of figure \ref{fig1}, which 
encompasses a larger number of sources. 
Moreover, to minimize possible effects introduced by the irregularity of the survey geometry, we restrict the analyses to the region defined by the coordinates $ 9^{\text{h}}20^{\text{m}} \leqslant \text{RA} \leqslant 15^{\text{h}}50^{\text{m}} $ and $ 0^{\circ} \leqslant \text{DEC} \leqslant 36^{\circ} $, as illustrated in the right panel of figure \ref{fig1}.
The final sample has 13144 HI sources of {\sc code 1}, whose redshift distribution is shown in 
figure~\ref{fig2}.

\section{Methodology}

This section describes the methodology adopted for our homogeneity analyses.
We present the scaled counts-in-spheres method, $\mathcal{N}(<r)$, and the equivalent quantity adapted to the 2D analyses we consider here, the scaled counts-in-caps (or spherical caps), $\mathcal{N}(<\theta)$, calculated through three different estimators, besides the related fractal correlation dimension, $\mathcal{D}_2$. 
Finally, we describe how they can be adapted to analyse the ALFALFA dataset, in addition to our criterium to determine the transition scale to homogeneity, $\theta_H$.

\subsection{The scaled counts-in-spheres: the 3D case}

To study the homogeneity of a dataset it is commonly adopted the counts-in-spheres estimator, $N(<r)$, which does not depend on hypotheses like homogeneity and completeness of the analysed sample. 
This quantity corresponds to the average number of objects encompassed by a sphere of a radius $r$, with $r_{\mbox{\footnotesize\rm min}} \,\leq\, r \,\leq\, r_{\mbox{\footnotesize\rm max}}$, centred around each object of the catalogue. 
Clearly, the choices for the values $r_{\mbox{\footnotesize\rm min}}$ and $r_{\mbox{\footnotesize\rm max}}$ to perform the numerical analyses depend on the features of the catalogue in scrutiny. 
For a 3D homogeneous distribution of objects one expects that 
\begin{equation}
N(< r) \propto r^{3}  \, . 
\end{equation}
For the general case of a fractal distribution of objects we have~\cite{fractal} 
\begin{equation}
N(< r) \propto r^{D_{2}}  \, ,
\end{equation}
where $D_{2}$ is termed the fractal correlation dimension (or just correlation dimension), defined as \begin{equation}
D_{2}(r) \,\equiv\, \dfrac{d \,ln \,N(<r)}{d \,ln \,r}. 
\end{equation}

However, it is not feasible to use this equation directly estimated from $N(<r)$, since the results can be 
biased by the survey geometry (in special, the boundary effects) and incompleteness of the 
sample~\cite{Laurent}. 
Aiming to reduce these effects it is introduced the estimator termed scaled counts-in-spheres, 
$\mathcal{N}(< r)$, defined as~\cite{Scrimgeour} 
\begin{equation}
\mathcal{N}(< r) \,\equiv\, \dfrac{N_{gal}(< r)}{N_{rand}(< r)} \, ,
\label{eqNnorm}
\end{equation}
where $N_{gal}(<r)$ is the average counting estimated for spheres centred at each source of the data 
catalogue and $N_{rand}(< r)$ is the same quantity calculated upon a random homogeneous sample, 
using as centres the coordinate positions of the data catalogue sources. 
Then, the correlation dimension in terms of $ \mathcal{N}(<r) $ is written as
\begin{equation}
\mathcal{D}_{2}(r) \equiv \dfrac{d \,ln \,\mathcal{N}(<r)}{d \,ln \,r} \,+\, 3 \, . 
\label{D2}
\end{equation}
As discussed in section 3.4 of ref.~\cite{Laurent}, for the scientific objective of our analyses, the estimator 
$\mathcal{D}_{2}$ is a more suitable estimator because, differently of $\mathcal{N}(<r)$, it is essentially 
uncorrelated between $r$ bins.

\subsection{The scaled counts-in-caps: the 2D case}

In this work we analyse the ALFALFA data projected on the sky, which requires the spheres of radius $r$ 
mentioned above, for the 3D case, to be replaced by spherical caps of angular radius 
$\theta$, so that Equation (\ref{D2}) must be accordingly modified.  
As detailed in ref. \cite{Rodrigo}, the $\mathcal{D}_{2}(r)$ estimator adapted to the 2D projection of the 
data sample on the sphere ${\cal S}^2$, $\mathcal{D}_{2}(\theta)$, 
is given by 
\begin{equation}\label{D2A}
\mathcal{D}_{2}(\theta) \,=\,  \dfrac{d \,\ln \mathcal{N}(<\theta)}{d \,\ln \theta} 
\,+\, \dfrac{\theta \, \sin\theta}{1-\cos\theta} \, , 
\end{equation}
where the counts-in-caps (or spherical caps) is given by
\begin{equation}
\mathcal{N}(<\theta) \equiv \frac{N_{gal}(<\theta)}{N_{rand}(<\theta)}.
\end{equation}
Note from Equation~(\ref{D2A}) that for a homogeneous angular distribution one has: 
$\mathcal{D}_{2}(\theta) \rightarrow \theta \, \sin\,\theta / (1 - \cos\,\theta)$, for sufficiently large $\theta$ value.

\subsubsection{Average, Centre, and Landy-Szalay estimators} \label{sec:estimators}

The literature provides some estimators for the $\mathcal{N}$ quantity (see, e.g., ref.~\cite{Rodrigo}). 
Here we employ three of them, namely, the {\em Average estimator} 
\begin{equation}\label{NA}
\mathcal{N}_{j}^{A}(<\theta)\equiv \dfrac{\frac{1}{n_{g}}\sum_{i=1}^{n_{g}}N^{i}_{gal}(<\theta)}{\frac{1}{n_{g}} 
\sum_{i=1}^{n_{g}}N_{rand}^{i,j}(<\theta)} \, , 
\end{equation}
the {\em Centre estimator}
\begin{equation}\label{NC}
\mathcal{N}_{j}^{C}(<\theta)\equiv \frac{1}{n_{g}}\sum_{i=1}^{n_{g}}\frac{N^{i}_{gal}(<\theta)}{N^{i,j}_{rand}(<\theta)} \, ,
\end{equation}    
where $n_{g}$ is the number of objects in the data sample, and the third estimator, that is based on the Landy-Szalay (LS) two-point angular correlation function~\cite{LS,deCarvalho}, 
\begin{equation}\label{w_theta}
\omega^{\mbox{\footnotesize LS}}_{j}(\theta) = \frac{DD(\theta) - 2 DR(\theta) 
+ RR(\theta)}{RR(\theta)} \, ,
\end{equation}
and for this called {\em LS estimator}
\begin{equation}\label{NLS}
\mathcal{N}_{j}^{LS}(<\theta)\equiv 1+ \frac{1}{1 - \cos\theta} \, \int_{0}^{\theta} \, 
\omega^{\mbox{\footnotesize LS}}_{j}(\theta') \, \sin \theta' \, d\theta'  \, , 
\end{equation}
calculated for the $j$th random catalogue. 
$DD(\theta)$ is defined as the number of pairs of galaxies in the data sample, for a given $\theta$, 
normalized to the total number of pairs; $RR(\theta)$ is the same quantity calculated for the random 
catalogue; and $DR(\theta)$ corresponds to the number of pairs between the data and random catalogues, 
normalized to the total number of pairs in the data and random catalogues.

Then, we use Equation \ref{D2A} to obtain the corresponding $\mathcal{D}^j_{2}(\theta)$ data points, for the $j$th random catalogue, in the range $1^\circ \leq \theta \leq 40^\circ$ and bins of $\Delta\theta = 1^\circ$. 
For this work we constructed 20 random homogeneous catalogues with the same geometry and number 
$n_g$ of objects as the ALFALFA catalogue. 
Therefore, the final $\mathcal{D}_{2}(\theta)$ quantity is then estimated from the arithmetic mean: $\mathcal{D}_{2}(\theta) = (1/20) \sum_{j = 1}^{20} \mathcal{D}^j_{2}(\theta)$, for each of the three estimators.
Notice that the integral in Equation \ref{NLS} is calculated over the best-fit curve of the $\omega^{\mbox{\footnotesize LS}}_{j}(\theta)$ data points, previously estimated using Equation \ref{w_theta} (see, e.g.,~\cite{Laurent,Ntelis}).

\subsection{The homogeneity scale criterium}

First we fit a model-independent polynomial to the $ \mathcal{D}_{2}(\theta) $ data points calculated as 
described in previous sections. Then, to determine the scale of transition to homogeneity, $ \theta_{H} $, 
we consider the 1\%-criterium commonly adopted in the 
literature~\cite{Scrimgeour,Ntelis,Laurent,Rodrigo}: we identify the scale at which the fitted curve 
reach 99\% of the $\mathcal{D}^{H}_{2}$ threshold value expected for a homogeneous distribution. 
For the case of a 2D (3D) Euclidean space this value is a constant,  $\mathcal{D}^{H}_{2} = 2$ 
($\mathcal{D}_{2}^{H} = 3$), while for a distribution on the 2D 
celestial sphere ${\cal S}^2$ it is a function of $\theta$ \cite{Rodrigo}, 
\begin{equation}\label{D2f}
\mathcal{D}_{2}^{H}(\theta)\equiv \dfrac{\theta~sin\theta}{1-cos\theta}. 
\end{equation}
Thus, the value $\theta_{\text{H}}$ is the scale where 
\begin{equation} \label{criterium}
\mathcal{D}_{2}(\theta_{H})=0.99~\mathcal{D}_{2}^{H}(\theta_{H}). 
\end{equation}
Despite the 1\%-criterium is arbitrary, it has positive attributes to have been widely adopted in the 
literature (for a complete discussion about the positive and negative attributes of several criteria 
see~\cite{Scrimgeour}); perhaps the most useful feature is its independence with the data sample in 
study, allowing an easy comparison among different analyses.

\section{Results}\label{results}

In this section, we present in detail the results we obtain applying the three estimators, 
\textit{Average}, \textit{Centre}, and \textit{LS}, to the ALFALFA catalogue. 
We also evaluate the effects that density fluctuations, given by structures like \textit{voids} and 
galaxy clusters, i.e. under- and over-densities, can introduce in the $\mathcal{D}_{2}$ and, 
consequently, in the estimates of the angular scale of transition to homogeneity. 
These analyses are specially important when the surveyed volume coverage is suitable to encompass 
such type of structures in the local universe, as is the case of the ALFALFA catalogue. 
We also perform consistency tests of the estimators, by using fractal catalogues, to verify if the use of 
$\mathcal{N}(<\theta)$ is biasing our results. 
Additionally, we generate and analyse realizations of segment Cox processes to confirm the suitability 
of the ALFALFA area for the current analyses.

\subsection{The angular scale of transition to homogeneity}\label{ss-trans}

The median redshift of the final sample selected from the ALFALFA catalogue is 
$z_{\rm m} \simeq 0.025$, for this we refer to it as a representative of the local universe. 
Our main results analysing this dataset can be seen in figure \ref{fig2}, for the {\em Average} and 
{\em Centre} estimators, and in the left panel of figure \ref{fig4}, for the {\em LS} estimator. 
As can be observed, the three estimators are in good agreement in the determination of the 
homogeneity angular scale, $\theta_{H}$, obtaining a value around $16^{\circ}$. 
In each case, this scale is measured through a polynomial fit for the 
$\mathcal{D}_{2}$ data points calculated for each of the 20 random catalogues individually (gray 
curves). 
We used a polynomial of order three in the case of the \textit{Average} and \textit{Centre} estimators, 
with an average {\it root mean square} error\footnote{The degree of the polynomial is chosen to be 
the one minimizing $E$, taking into account the Akaike Information Criterium to penalize the number 
of free parameters of the fit.} 
of $E = 0.0064$ and $0.0078$, respectively, and of order five for the \textit{LS} estimator, with 
$E = 0.0012$. 
In the estimates showed in table~\ref{table1}, the error bars correspond to the 68\% confidence 
level obtained from a \textit{bootstrap resampling} of the 20 values of $\theta_H$ estimated using 
each of the random catalogues.

From the analyses presented in refs \cite{Alonso14,Alonso15}, performed using simulated catalogues 
generated with bias $b = 1$ and redshift ranges of thickness comparable to ours, we extrapolate the 
$\theta_H$ estimated there to the redshift range we are considering, obtaining $32.46^\circ$ for the 
expected value of the homogeneity scale. 
However, according to ref.~\cite{Basilakos}, extragalactic HI line sources have an anti-bias $b$ with 
respect to the matter fluctuations ($b = 1$), namely, a factor that varies between  0.48 -- 0.68, depending 
on the richness of the sample. 
Therefore, the expected value for the homogeneity scale, for the specific case of the data we are 
analysing, is estimated to be $15.58^{\circ} \lesssim \theta_H \lesssim 22.07^{\circ}$ (table \ref{table1}).

\begin{figure}[h]
\mbox{\hspace{-0.4cm}
\includegraphics[width=8.cm, height=5.8cm]{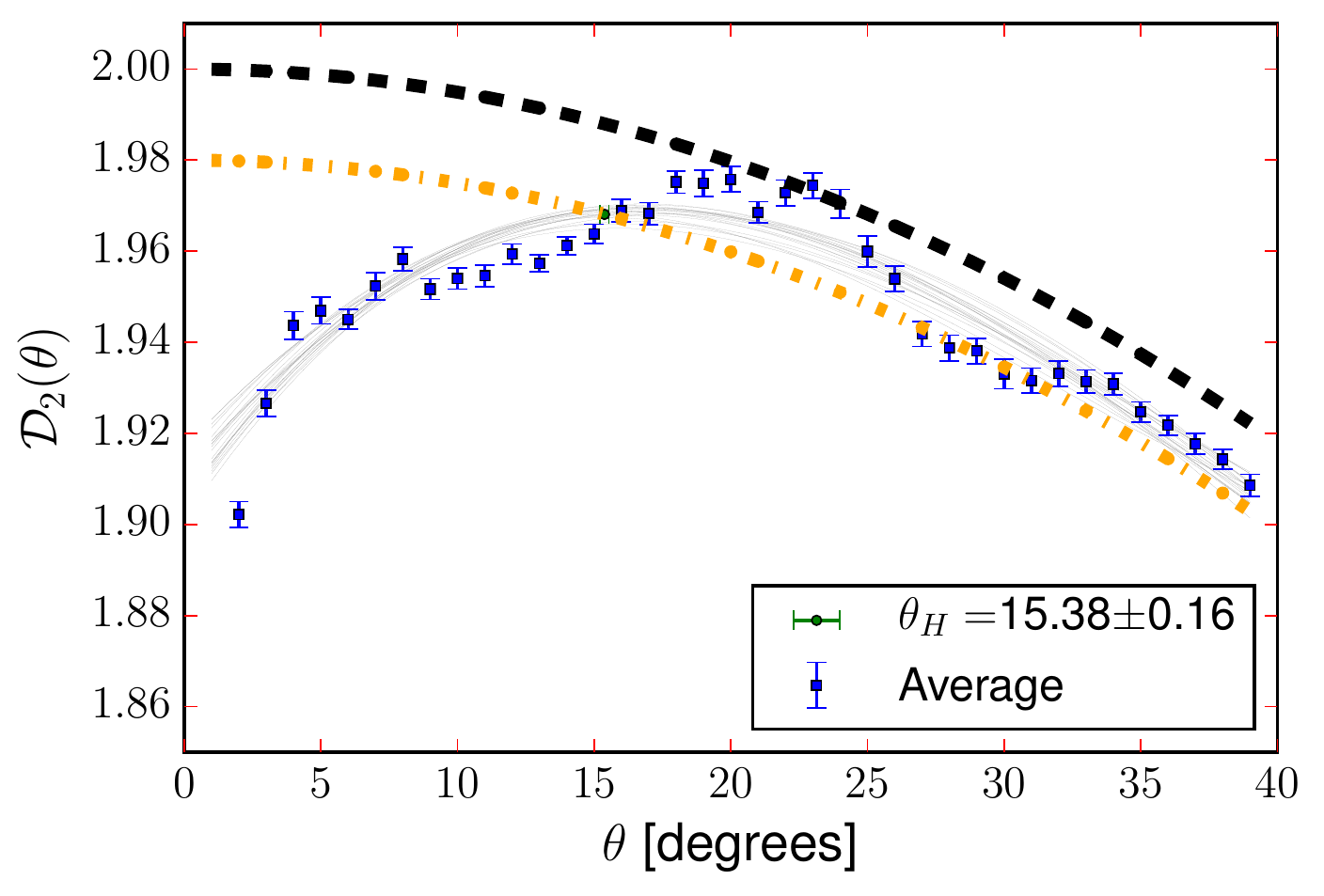}
\hspace{-0.4cm}
\includegraphics[width=8.cm, height=5.8cm]{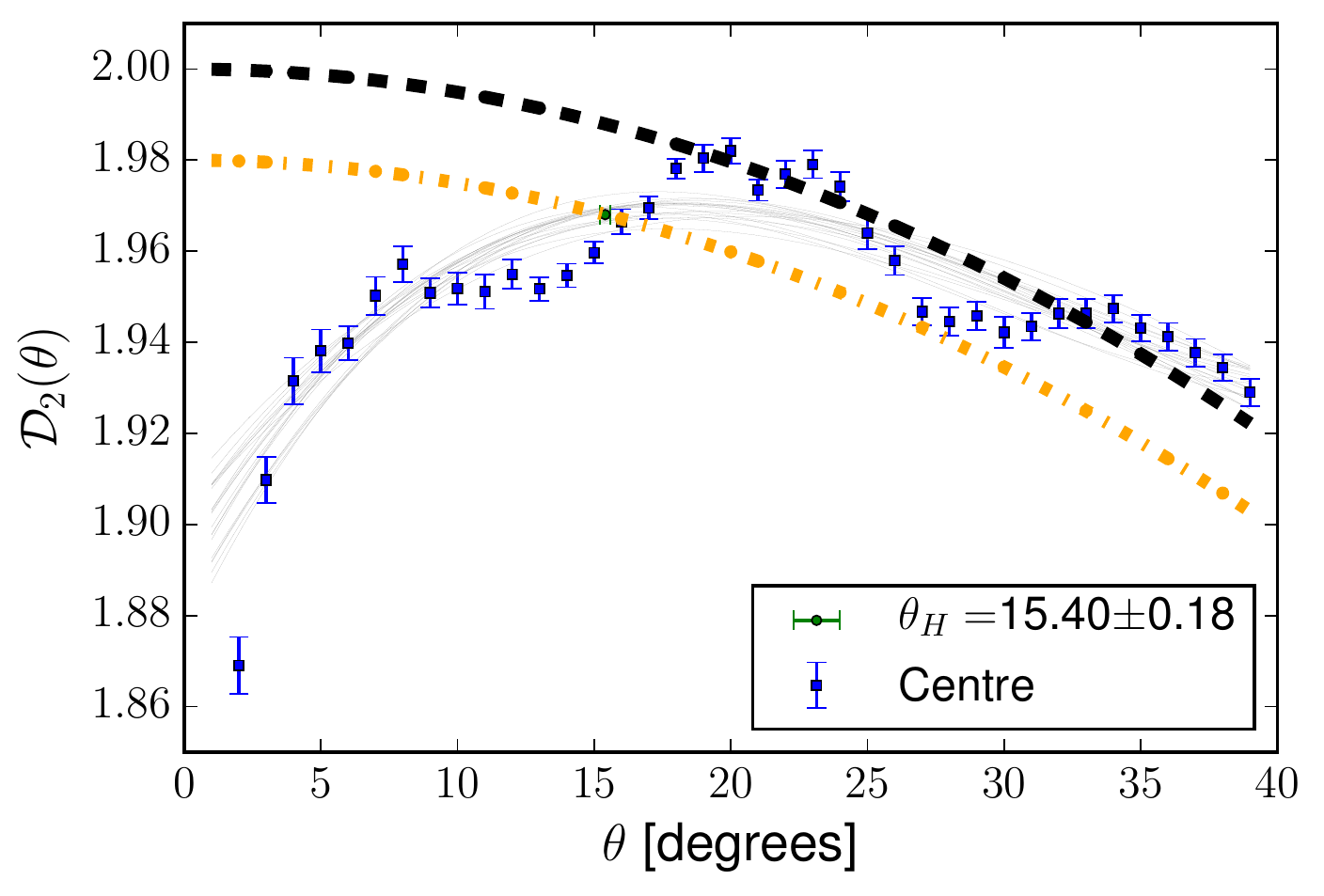}
}
\caption{
{\bf Left}: $\mathcal{D}_{2}(\theta)$ data points (blue dots) obtained from Equation \ref{D2A} for the 
{\em  Average} estimator (Equation \ref{NA}) in the range $1^{\circ} \leqslant \theta \leqslant 40^{\circ}$, 
with bins size $\Delta \theta = 1^{\circ}$. 
The corresponding error bars were obtained from the standard deviation from the 20 random catalogues. 
The black dashed line represents the threshold value for $\mathcal{D}_{2}(\theta) $ when the distribution 
is considered homogeneous, i.e., the $\mathcal{D}^H_{2}(\theta)$ curve given by Equation \ref{D2f}. 
The orange dot-dashed line represents the 1\% below this threshold (the 1\%-criterium given by 
Equation \ref{criterium}). 
The gray curves correspond to polynomial fits of order three for the $\mathcal{D}_{2}(\theta)$ points 
measured using each one of the 20 random catalogues, whose intersection with the orange dot-dashed 
line determines the angular scale of transition to homogeneity, $\theta_H$. 
The green dot is the mean over these 20 values and its error bar is obtained using the bootstrap method 
(see subsection~\ref{ss-trans} for details). 
{\bf Right}: Similar to the information displayed in the left panel, but now the data analysis was done with 
the {\em Centre} estimator (Equation \ref{NC}).
}
\label{fig3}
\end{figure}

\begin{figure}
\centering
\mbox{\hspace{-.4cm}
\includegraphics[width=8.cm, height=5.8cm]{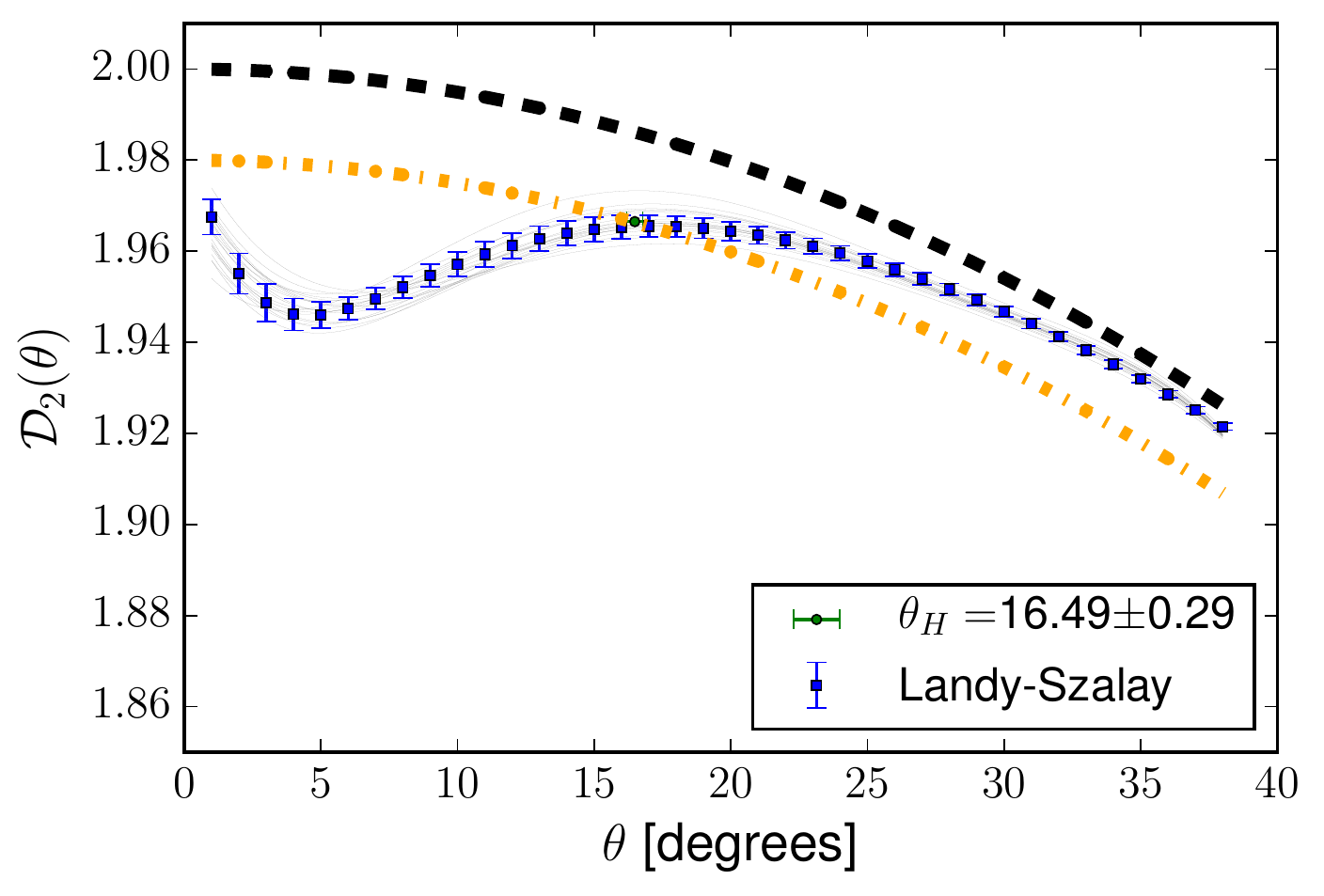}
\hspace{-0.4cm}
\includegraphics[width=8.cm, height=5.8cm]{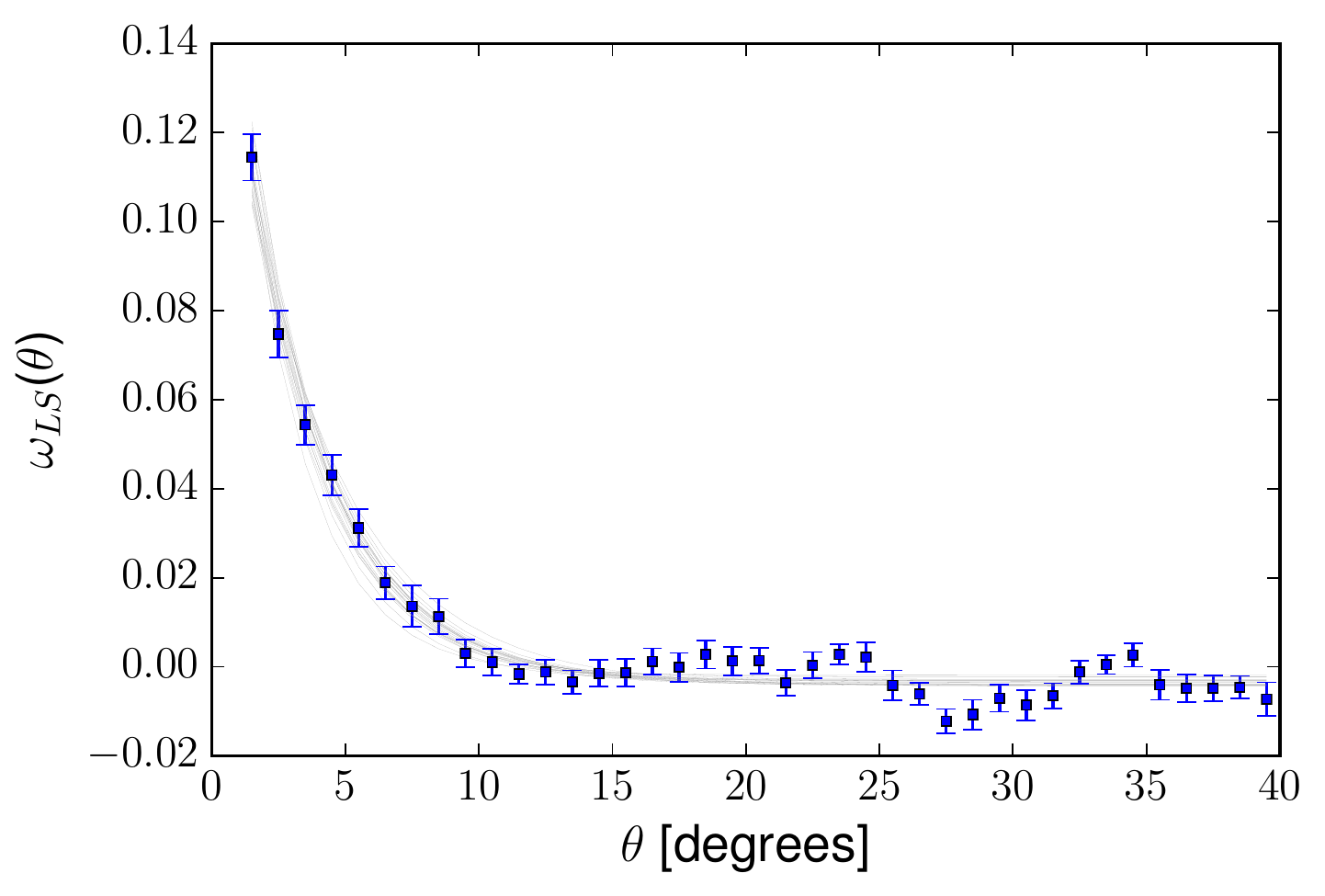}
}
\caption{
{\bf Left}: Similar to the information displayed in figure~\ref{fig3} but for the analyses done with the 
{\em LS} estimator (Equation \ref{NLS}). 
{\bf Right}: Average of the {\em LS} two-point angular correlation function, $\omega_j^{LS}$, calculated 
through Equation \ref{w_theta}, for the 20 random catalogues, 
$\omega^{LS} = \langle \omega_j^{LS} \rangle$, for $j = 1, ..., 20$ (blue dots). 
The error bars correspond to the 1$\sigma$ dispersion over these 20 estimates. 
The gray curves are the best-fit of the $\omega_j^{LS}$ data points, used to calculate 
$\mathcal{D}^j_{2}(\theta)$ for the $j$th random catalogue (see section \ref{sec:estimators} for details).
}
\label{fig4}
\end{figure}

\begin{table}[h]
\centering
\begin{tabular}{| c | c | c | c | c |}
\hline
                              & {\em Average} estimator & {\em Centre} estimator & {\em LS} estimator 
& expected~\cite{Alonso14,Alonso15}  \\
\hline
\,\,$\theta_{\text{H}}$\,\, &  $15.38^{\circ} \pm 0.16^{\circ}$ & 15.40$^{\circ} \pm 0.18^{\circ}$ & 
16.49$^{\circ}\pm0.29^{\circ}$ &  $[15.58^{\circ}, 22.07^{\circ}]$  \\
\hline
\multicolumn{5}{p{2.5cm}}{\,} 
\end{tabular}
\caption{In this table we summarize our findings, that is, the angular scale of transition to homogeneity as measured by our three estimators.
We also show, in the last column, the value expected according to simulations, done for matter fluctuations ($b=1$) \cite{Alonso14,Alonso15}, that we extrapolated to our sample considering the ALFALFA redshift range and the appropriate bias factor \cite{Basilakos}.  
} \label{table1}
\end{table}

\subsection{Effects introduced by density fluctuations} \label{clusters}

First of all, it is worth to notice the features clearly appearing in the $\mathcal{D} _{2}(\theta)$ curves obtained from the {\em Average} and {\em Centre} estimators (figure \ref{fig3}), presenting successive data points bellow and above the $\mathcal{D}^H_{2}(\theta)$ threshold expected for a homogeneous distribution (black dashed line given by Equation \ref{D2f}). 
This is still confirmed by a similar behaviour of the $\omega^{LS}$ data points around the best-fit curves 
(gray lines) in the right of figure~\ref{fig4}, a signature likely associated to density fluctuations on the 
sample\footnote{Notice that these features do not appear in the $\mathcal{D}_{2}$ obtained with the {\em LS} estimator because we use the best-fit curve over the $\omega_j^{LS}$ data points to estimate $\mathcal{N}_{j}^{LS}(<\theta)$, from which we obtain the corresponding $\mathcal{D}_{2}^j$ and mean $\mathcal{D}_{2}$.}.  
Such behaviour seems not only to indicate the presence of under- or over-dense regions in the data sample, but also, and more interestingly, to furnish information about the extension and how under- or over-dense could be such regions.
Specifically, one can identify features appearing in the scale intervals of $\sim 8^{\circ}$ -- $16^{\circ}$ 
(this one only in the {\em Average} and {\em Centre} $\mathcal{D}_{2}$) and $\sim 25^{\circ}$ -- $33^{\circ}$, 
which could be associated to under-densities, while the one in scale range $\sim \,20^{\circ} - 25^{\circ}$ 
seems to be influenced by the presence of an over-density.  
Therefore, bellow we present results of additional analyses performed not only to verify these statements, 
but also to evaluate how under- and over-densities appearing in the data sample can affect our analyses.

\subsubsection{Over-densities}

\begin{figure}[!h]
\centering
\mbox{\hspace{-0.4cm}
\includegraphics[width=8.cm, height=5.8cm]{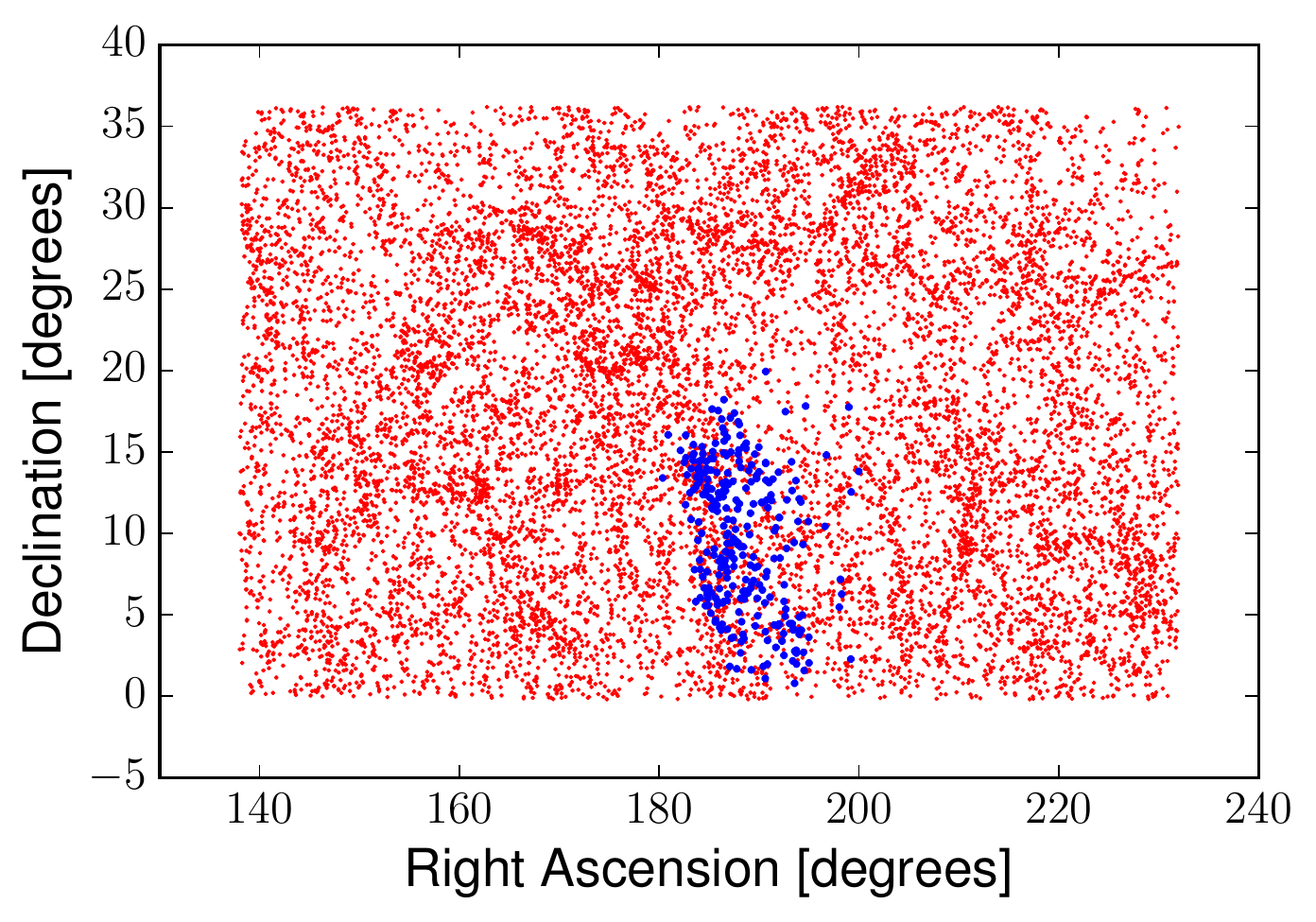}
\hspace{-0.4cm}
\includegraphics[width=8.cm, height=5.65cm]{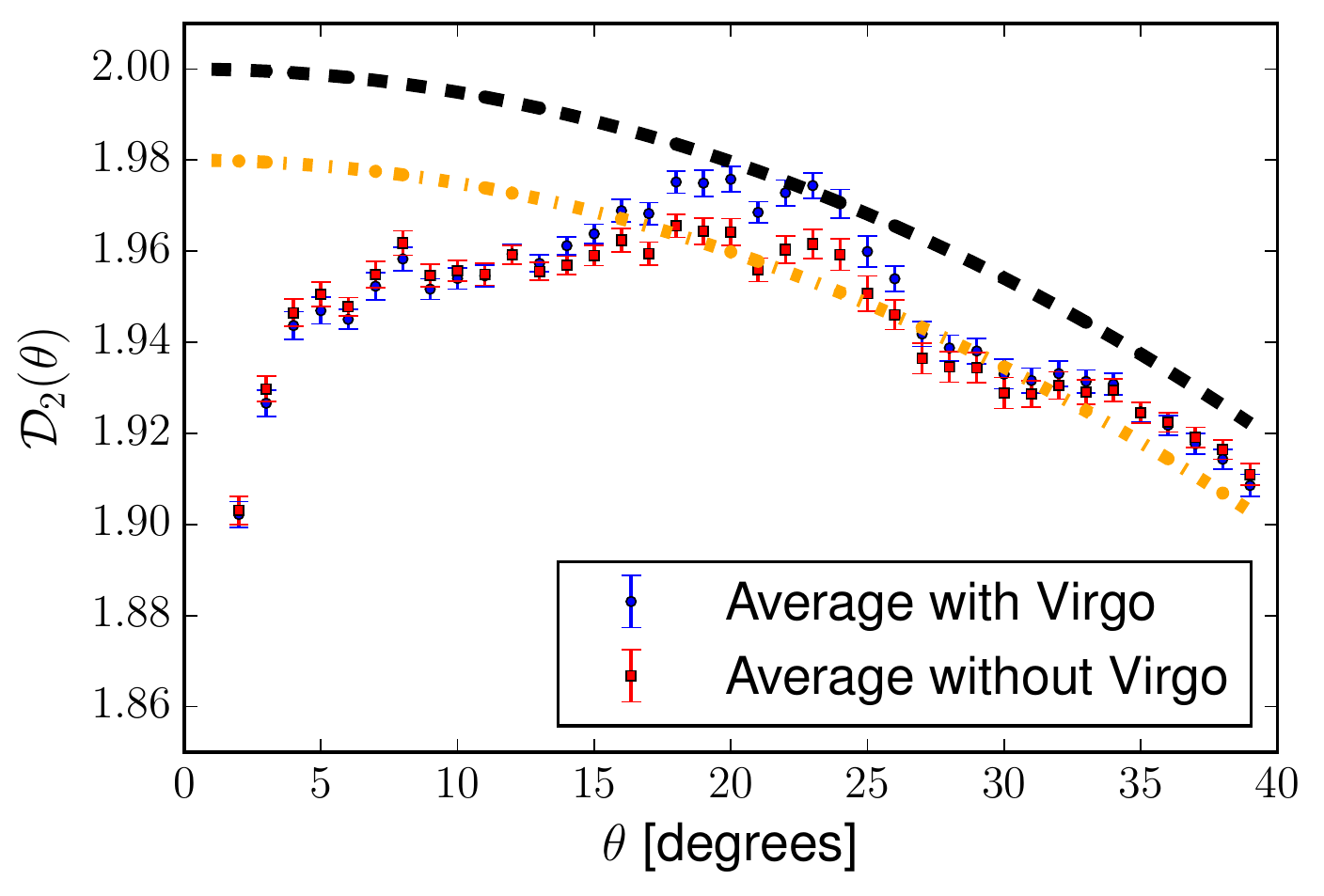}
}
\caption{
{\bf Left}: Plot of the selected ALFALFA sample highlighting in blue color the 287 objects in between 
$16 - 18$ Mpc in distance, $ 0^{\circ}\leqslant\text{DEC}\leqslant20^{\circ} $ in declination, and 
$12^{\text{h}}\leqslant\text{RA}\leqslant13^{\text{h}}30^{\text{m}} $ in right ascension, possibly belonging 
to the Virgo cluster. 
{\bf Right}: For comparison, we show the $\mathcal{D}_{2}(\theta)$ estimates 
obtained with the {\em Average} estimator when analysing the data sample with and without the objects highlighted in blue in the left panel, i.e., the Virgo cluster. 
Again the black dashed and orange dot-dashed line represent, respectively, the $\mathcal{D}^H_{2}$ threshold for a homogeneous sample (Equation \ref{D2f}) and the 1\% bellow it, given by Equation \ref{criterium}.
}
\label{fig5}
\end{figure}

In this section we investigate in detail the possible presence of an over-dense region suggested by the fact 
that the $\mathcal{D}_{2}(\theta)$ data points overtake the threshold curve $\mathcal{D}_{2}^H(\theta)$ 
(given by the black dashed line in figure~\ref{fig3}; see also Equation~\ref{D2f}). 
To elucidate if a clustering of objects in a small region can produce this effect we first identify the candidate 
objects, remove them from the catalogue in study, and then repeat our analyses. 
The objects we chose to remove belong to the Virgo cluster, a known structure appearing in the data 
sample we analyse.
In the principal reference of the ALFALFA survey~\cite{haynes}, the objects belonging to the Virgo cluster 
are identified as mainly standing at distances between $16 - 18$ Mpc. 
Moreover, we use the recent catalogue of sources belonging to the Virgo cluster to identify the angular 
region encompassing most of them, namely, $0^{\circ}\leqslant\text{DEC}\leqslant20^{\circ}$ and 
$12^{\text{h}}\leqslant\text{RA}\leqslant13^{\text{h}}30^{\text{m}} $ \cite{kim2014}.
For illustrative purposes, in the left panel of figure~\ref{fig5} we highlight with a different color the objects 
belonging to this region, from where it is visually noticeable that they appear clustered. 
Finally, using the {\em Average} estimator we show at the right panel of figure~\ref{fig5} a comparison 
among the analyses of the ALFALFA sample with and without the presence of the Virgo cluster. 
One can clearly observe from this graph that without these objects the $\mathcal{D}_{2}(\theta)$ data 
points remain bellow the threshold curve; with a similar behaviour from the {\em Centre} estimator. 
Moreover, notice that the difference between the two analyses, with and without this structure, is restricted 
to scales around $\theta \sim 20^\circ$, indicating a possible relation with the cluster dimension. 
Therefore, this confirms our preliminary suspicion that, indeed, the Virgo cluster is an over-density that manifests itself by overtaking the $\mathcal{D}^H_{2}$ curve at the scale $20^{\circ} - 25^{\circ}$.

\subsubsection{Under-densities}\label{voids}

\begin{figure}[!h]
	\centering
	\hspace*{-0.6cm}
	\begin{minipage}{.52\textwidth}
		\centering
		\includegraphics[width=1\linewidth]{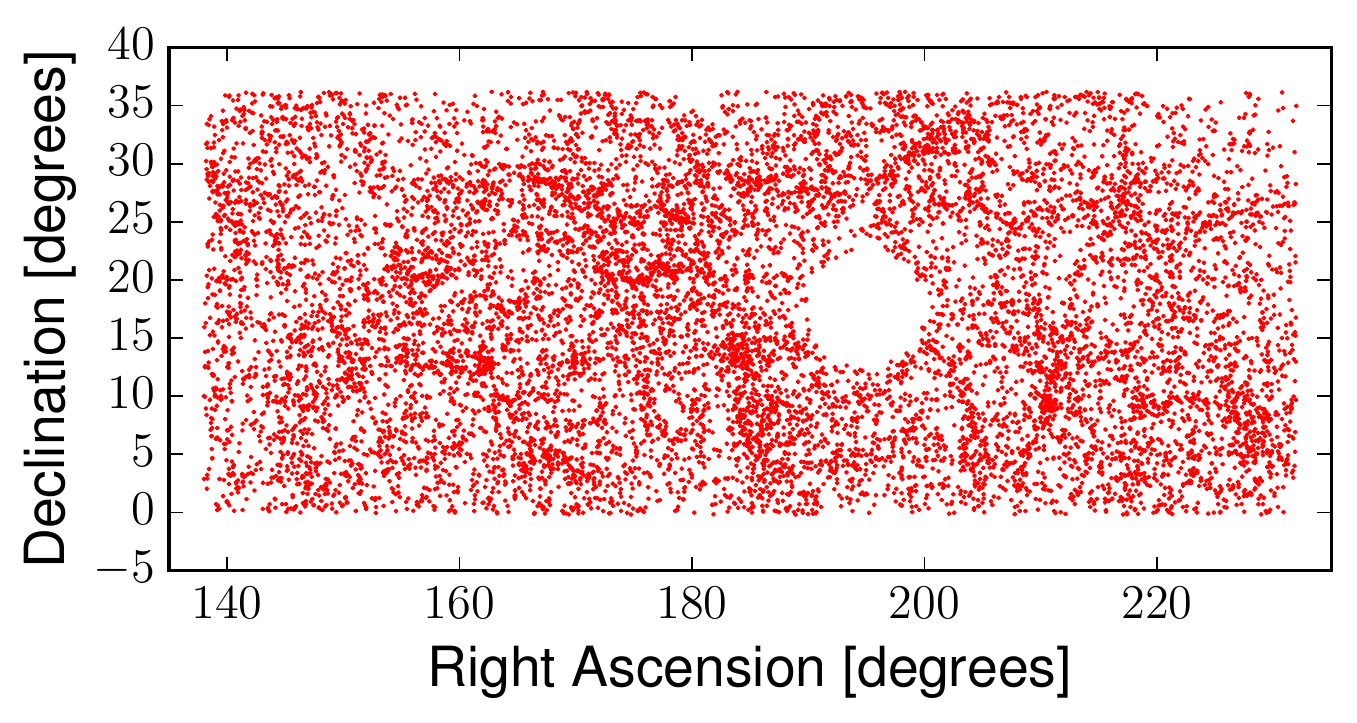}
	\end{minipage}%
	\begin{minipage}{.52\textwidth}
		\centering
		\includegraphics[width=1\linewidth]{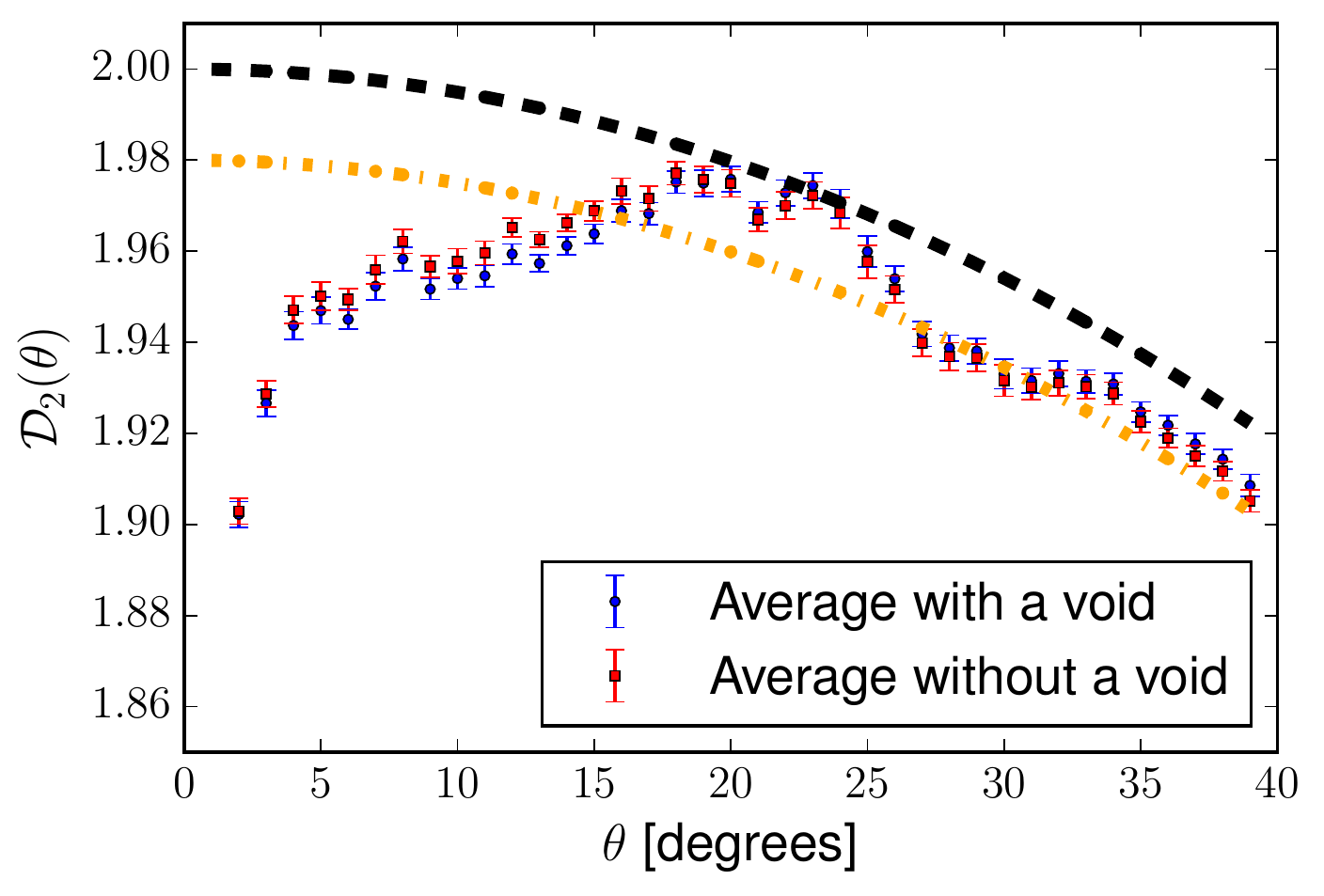}
	\end{minipage}
	\caption{{\bf Left}: ALFALFA sample after removing an under-dense area. The masked region correspond to a circle of $5^{\circ}$ radius (i.e., an object of size $10^{\circ}$) 
centred at RA, DEC = (193$^\circ$, 17.5$^\circ$).
		{\bf Right}: $\mathcal{D}_{2}(\theta)$ estimated applying the {\em Average} estimator to the ALFALFA sample before (blue dots) and after (red squares) masking an under-dense area.}
	\label{fig6}
\end{figure}

As mentioned above, our $\mathcal{D}_{2}(\theta)$ analyses suggest the presence of under-densities at 
the scale intervals $8^{\circ} - 16^{\circ}$ and $25^{\circ} - 33^{\circ}$. 
In order to verify this, we perform a test analogous to the one in the previous section, 
but removing (masking) the under-dense region, a disk of radius $5^{\circ}$ at RA =  $13^{\text{h}}$ 
($193^{\circ}$) and DEC = $17.5^{\circ}$ (left panel of figure~\ref{fig6}). 
Such area was identified by using a void-finder like code, which perform a scan in the whole projected 
data sample looking for the region of lowest number density. 

The right panel of figure~\ref{fig6} shows the $\mathcal{D}_{2}(\theta)$ data points 
obtained by applying the \textit{Average} estimator to the masked ALFALFA sample (red squares), 
where we also include the results from the non-masked sample (blue dots). 
From the comparison among them we conclude that, as expected, the presence of under-dense 
regions in the sample tends to decrease the $\mathcal{D}_{2}(\theta)$ values corresponding to 
the size of that regions. 
This can be seen from the amplitude of these data points around $8^{\circ} - 16^{\circ}$, which 
increases when excluding the under-dense region from the analysis, the opposite behaviour to 
that observed when we removed the Virgo members. 
Therefore, this confirms our expectation that the presence of both, over- and under-densities, have 
a definite influence in the behaviour of the $\mathcal{D}_{2}(\theta)$ function and impact the scale 
transition to homogeneity.

\subsection{Performance test with simulated fractal samples}\label{fractals}

In this section we use fractal realizations, constructed according to a predefined $\mathcal{D}_{2}$ value, to look for possible bias introduced in $\mathcal{N}(<r)$ by the methodology considered here, in special, due to boundary effects.
For this test we consider 100 realizations, with a mean of $\sim 22000$ points distributed on the same geometry as the selected ALFALFA data but on a 2D Euclidean plane ${\cal R}^2$. 
Then we compute $\mathcal{N}(< r)$ and $\mathcal{D}_{2}(r)$ using disks, where distances $r$ (in arbitrary units) are calculated with the Euclidean geometry. 

Each realization is a self-similar fractal sample generated using a modified 
$\beta$-model~\cite{fractal}. 
First, consider a square of side $L$, and divide it in $M = n^{2}$ squares of side $L/n$. 
Then, impose a survival probability $p$ to each square at each iteration, and repeat this process $k$ times. 
In the limit of infinite iterations we would have 
\begin{equation}\label{D2p}
D_{2} \,=\, \lim\limits_{k \,\rightarrow\infty}\, \dfrac{log(pM)^{k}}{log \, n^{k}}
          \,=\, \dfrac{log(p M)}{log\, n}  \, .
\end{equation} 
The value of $D_2$ defines the probability $p$ through the Equation~\ref{D2p}. 
For our purposes to construct realizations with fractal dimension $D_2 = 1.9$, it is enough to perform $k = 7$ iterations in order to get a sample with approximately the same number of points as objects in the selected data.
What we expect to recover in this analysis is therefore $D_2\simeq 1.9$, for large distances. 

\begin{figure}[!h]
\centering
\mbox{\hspace{-0.7cm}
\includegraphics[width=8.6cm, height=6.3cm]{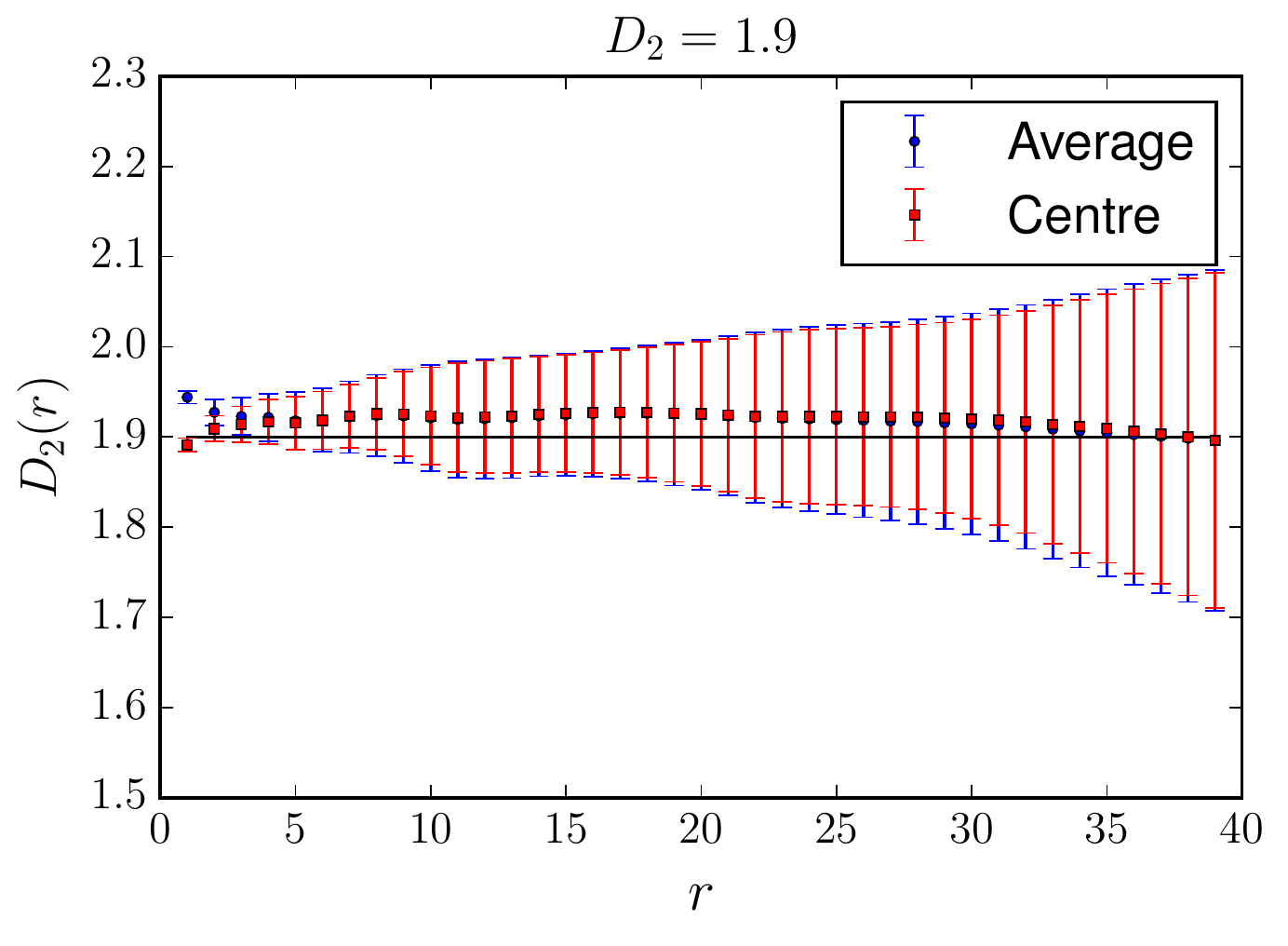}
\hspace{-0.8cm}
}
\vspace{-0.5cm}
\caption{
Average $\mathcal{D}_{2}(r)$ estimated from 100 fractal realizations, generated with an approximate 
number of points as objects in the ALFALFA catalogue and distributed in the same geometry as in the left 
panel of figure~\ref{fig6}. 
The error bars represent the 1$\sigma$ deviation from these samples. 
The black line corresponds to the input $\mathcal{D}_{2} = 1.9$ value. 
}
\label{fig7}
\end{figure}

We applied the {\em Average} and {\em Centre} estimators to the 100 fractal realizations, using 10 random 
homogeneous catalogues in each evaluation of $D_2$. 
The random catalogues were produced considering homogeneously distributed data in the same geometry 
and with the mean number of points as each of the 100 fractal samples (strictly speaking, since the number 
of points can vary from one fractal sample to another, different samples require random catalogues with 
different number of points). 
Our results can be seen in figure~\ref{fig7}, where we satisfactorily recover the input fractal dimension, obtaining the mean values $ \bar{\mathcal{D}_{2}} = 1.91818 $ for the {\em Average} estimator and $ \bar{\mathcal{D}_{2}} = 1.91824 $ for the {\em Centre} estimator. The error bars are the standard deviation from the 100 realizations. We also test for others fractal dimensions, namely 1.8 and 1.85, recovering values close to the input ones.

\subsection{Robustness tests with the segment Cox process}\label{Cox}

Additionally to the previous tests, performed to evaluate the robustness of our results and 
enhance our comprehension of the features observed, this section presents one additional 
test whose aim is to evaluate the influence of the observational constraint considering 
the size of the area selected with ALFALFA data. 
For this, we use the {\it segment Cox point process}, i.e., artificial catalogues generated 
given a characteristic length scale $l$~\cite{martinez98,mj99}. 
Such samples are generated by randomly placing segments of length $l$ in a box of side 
$L$, with a mean number of segments per unit volume $\lambda_s$, i.e., a length density 
of segments $L_V = \lambda_{s} \, l$, and then randomly distributing points on these 
segments. 
The intensity of the point process, i.e., the number density of points in the volume, is given 
by $\lambda = \lambda_{l} L_{V} = \lambda_{l} \lambda_{s} l$, where $\lambda_{l}$ is the average number of points per unit length of a segment.

\begin{figure}[h]
	\mbox{\hspace{-0.4cm}
		\includegraphics[width=8.cm, height=5.8cm]{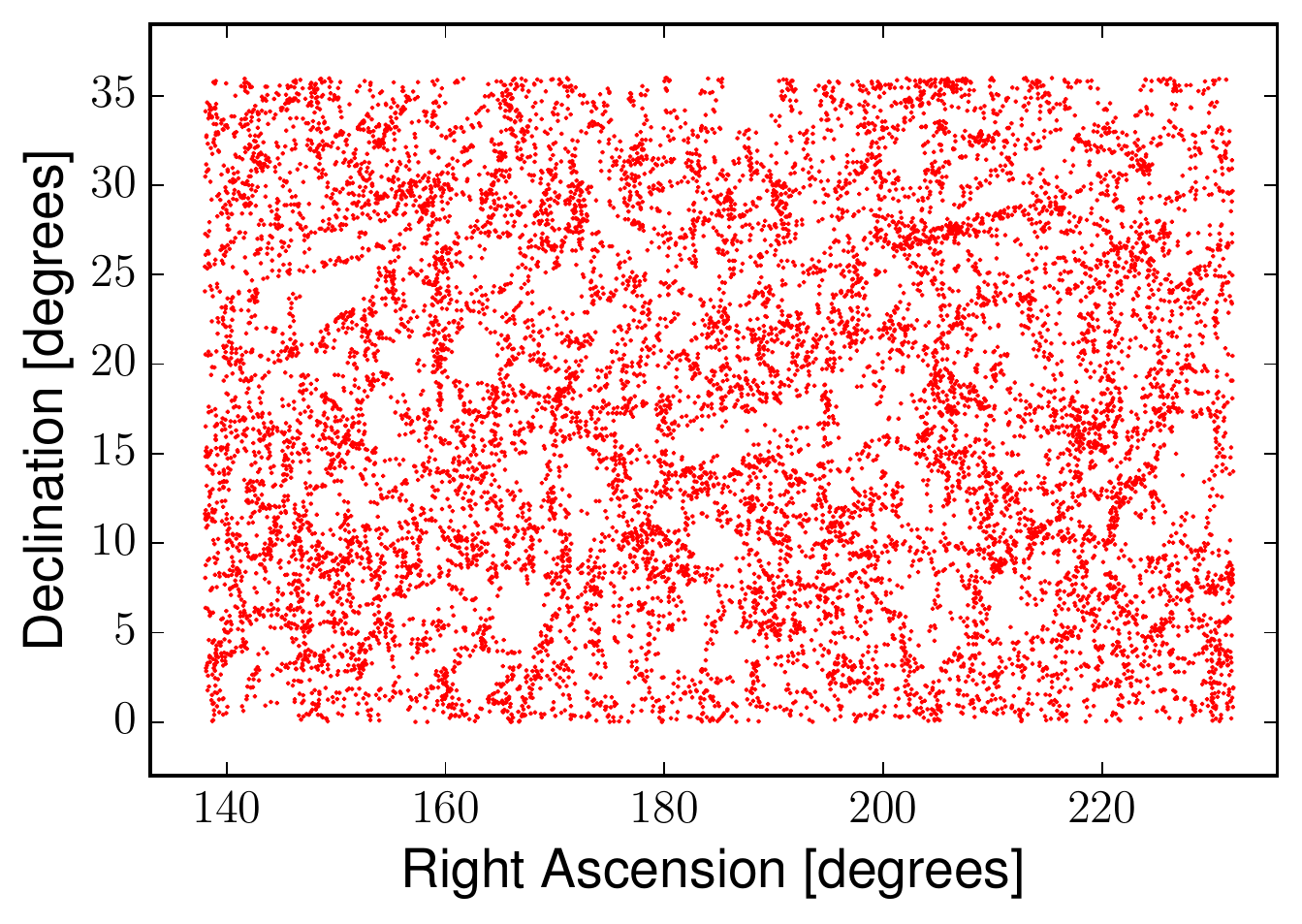}
		\hspace{-0.3cm}
		\includegraphics[width=8.cm, height=5.8cm]{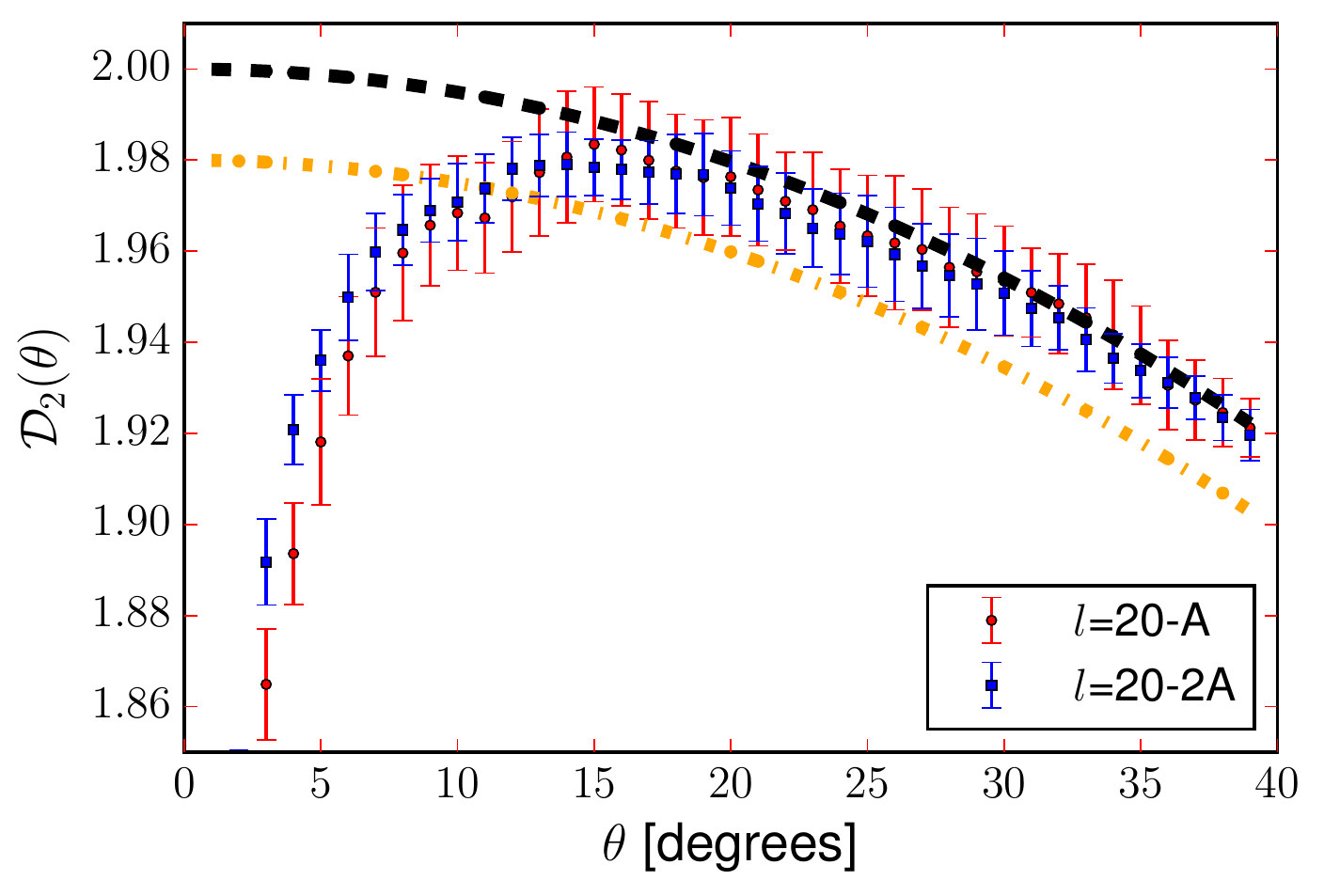}
	}
	\caption{
		{\bf Left}: Illustrative example of a segment Cox process for $l=20$, with the same 
number density of objects and projected area A of the ALFALFA sample, after performing a 
Gaussian shift of the data positions with $\sigma=0.5$. 
		{\bf Right}: Correlation dimension $\mathcal{D}_{2}(\theta)$ averaged over 20 segment 
Cox processes constructed with $l = 20$ Mpc on the area $A$ (red dots) and other 20 on area $2A$ 
(blue squares). 
The error bars correspond to one standard deviation of the data from these simulations. 
These analyses were done using the \textit{Average} estimator.
	}
	\label{fig8}
\end{figure}

We generate 20 segment Cox processes for different values 
of $l$, namely, $l=$ 10 and 20 Mpc, fixing $\lambda_{l}=1$ and varying $\lambda_{s}$ in 
such a way to achieve a number density close to the ALFALFA projected sample. 
These samples were constructed considering a box of volume $2.16\times10^{8}$ Mpc$^{3}$, 
where the observer is at the center. 
In this way the projection of the simulated catalogue results in a distribution of objects in the 
whole 2D sphere, so that we are free to take a region of any area needed. 
We construct and analyse segment Cox processes of two projected areas: $A$, 
which is the area of the ALFALFA sample, and $2A$, twice its area; the second one 
corresponds to the region delimited by the coordinates 
$0^{\circ} \leqslant \text{DEC} \leqslant 72^{\circ}$ and 
$138^{\circ}\leqslant \text{RA} \leqslant 232^{\circ}$. 
Before projecting the sample, we randomly shift the points by using a Gaussian distribution 
of $\sigma=0.5$~\cite{martinez_saar_2002}, 
smearing out the strong clustering of the segments, which helps to diminish the noise in 
the resulting $\mathcal{D}_{2}$. 
An example of a segment Cox process is exhibited in the left panel of 
figure~\ref{fig8}.

Each one of the 20 segment Cox processes, for given $l$ (10 or 20 Mpc) and area ($A$ or 
$2A$), was analysed following the same procedure used to analyse the ALFALFA data 
sample, as well as the 1\% criterium to determine the transition scale. 
Moreover, each sample was analysed using the same 20 random catalogues used previously 
(section~\ref{ss-trans}), i.e., one has 20 values of $\theta_{H}$ for each simulated catalogue, 
leading to a total of $20 \times 20 = 400$ values of $\theta_{H}$ for given $l$ and area. 
Table \ref{table2} summarizes the results obtained from the \textit{Average} estimator, 
presenting in the third and fifth columns the mean and standard deviation of each set of 
$400$ values, for the cases where the area is equal to $A$ and $2A$, respectively. 
It is evident the very good agreement between the $\theta_{H}$ estimates analysing 
different areas, within their error bars, showing smaller values for $2A$ due to the better statistics 
for a larger area.

A comparison among the results from different areas can also be seen from the right panel 
of figure~\ref{fig8}, showing the general behaviour of the $\mathcal{D}_{2}$ points averaged 
over 20 segment Cox processes constructed with $l = 20$ Mpc for $A$ (red dots) and $2A$ 
(blue squares). 
One observes in this plot the same asymptotic behaviour and equal transition scale to 
homogeneity, in both cases using the 1\% criterium. 
Therefore, the analyses of different segment Cox processes let us to conclude that the 
projected area of the ALFALFA sample is suitable for the analyses we perform here. 

\begingroup
\setlength{\tabcolsep}{10pt}             
\renewcommand{\arraystretch}{1.2} 
\begin{table}[h]
	\centering
	\begin{tabular}{| c || c | c || c | c | c |}
		\hline
 $l$ [Mpc] & $ \lambda^{A}_{s} $ & $ \theta^{A}_{H} $ & $ \lambda^{2A}_{s} $ & $ \theta^{2A}_{H} $ \\
	\hline
		10 & $ 2.31\times 10^{-4} $ & $9.79^{\circ} \pm 0.63^{\circ}$ & $2.78\times 10^{-4}$ & $9.38^{\circ} \pm 0.37^{\circ}$ \\
		\hline
		20 & $ 1.48\times 10^{-4} $ & $11.13^{\circ} \pm 1.33^{\circ}$ & $ 1.16\times 10^{-4} $ & $10.62^{\circ} \pm 0.61^{\circ}$ \\
		\hline
	\end{tabular}
	\caption{Results from applying the {\it Average} estimator to segment Cox processes 
for two different $l$ values, 10 and 20 Mpc, and two projected areas, $A$ and $2A$. 
The displayed $\theta_{H}$ values and error bars correspond to the mean and one standard 
deviation of the total of 400 values estimated from the 20 simulations of each type (20 
$\theta_{H}$ values for each simulation; see the text for details). 
In all the cases we used $\lambda_l = 1.0$.}
	\label{table2}
\end{table}
\endgroup

\section{Conclusions}

The recent progress in surveying tracers of the large scale structure has substantially 
increased the volume of the observed universe. 
This facility has allowed the exploration of the CP, in special the analyses of the homogeneity 
of the Universe, which require a high density of cosmic objects. 
Although we are not capable to prove its validity, one can still perform consistency tests of 
such property. 
In this work we used the scaled counts-in-caps and the fractal correlation dimension to 
measure the angular scale corresponding to the transition to homogeneity in the local 
universe ($0 < z < 0.06$), using the recently released ALFALFA catalogue. 
The methodology adopted, besides it has been extensively employed in the literature, 
exhibits the advantage that it does not rely on the geometric details or incompleteness of the 
data survey. 

Our angular scale analyses, in the 2D sky projected data from ALFALFA, reveal a transition 
to homogeneity that appears robust because: 
(i) we have performed analyses considering three estimators obtaining, basically, the 
same transition scale (see section~\ref{results}); 
(ii) interesting features appearing at several scales in the 
$D_2(\theta)$ data points, are --after detailed analyses-- associated with under- and over-densities present in the dataset (see sections~\ref{clusters} and~\ref{voids}); 
(iii) our procedures were successfully tested with simulated catalogues, 
including fractal realizations (see section~\ref{fractals}) and segment Cox processes 
(see section~\ref{Cox}).

Regarding the use of three estimators, it is worth to mention that one estimator can reveal aspects that other estimator ignores, for this the convenience of appreciate the features revealed by diverse estimators. 
In fact, as observed in section~\ref{results}, the {\em Average} and {\em Centre} estimators 
reveal the presence of under- and over-dense regions through features appearing in the 
$D_2(\theta)$ data points, while the {\em LS} estimator expose them directly in the 2-point 
correlation functions, $\omega^{LS}$. 
And with respect to possible observational effects due to matter clustering or voidness in 
the survey, the way we analyse them is with the aid of simulations that help us to verify their 
plausible presence through the signatures they leave in the estimators, even when they are 
not so large neither visually observed in the surveyed area. 
As a consequence of such detailed analyses, we show that clusters and voids can affect 
the measurement of the transition scale to homogeneity. 
Additionally, we also investigate if the size of the projected 2D area of the ALFALFA selected 
sample can influence the measurement of the transition scale to homogeneity, $\theta_H$. 
By applying the \textit{Average} estimator to segment Cox processes of different projected 
areas, one with the same area of the ALFALFA and the other with twice its area, we achieved 
results consistent one with each other within their 1$\sigma$ error bars 
(see Table~\ref{table2} and figure~\ref{fig8}). 
This confirms the suitability of the ALFALFA surveyed area for the current analyses.

Other interesting issues regard observational systematics, like instrumental effects, masks, 
possible contaminations, survey geometry, etc., which can be source of errors or 
uncertainties. 
For this, the ALFALFA team made efforts to correct or minimize them, resulting in a dataset 
of excellent quality detection in HI, objects termed as {\sc Code} 1, considered here in our analyses. 
In addition, in the current release the survey is 100\% complete and has a very good ratio between 
surveyed area and detected number of objects.

Our main result is presented in table~\ref{table1}. 
There we show the estimate obtained for the angular scale of transition to homogeneity, 
$\theta_{H} \simeq 16^{\circ}$, when analysing the local universe data given by the 2D 
projection of the final selected sample of the ALFALFA catalogue (see section~\ref{sec2}). 
This estimate is in agreement with the value expected according to simulations 
assuming a fiducial $\Lambda$CDM cosmology~\cite{Alonso14,Alonso15}. 
For this we confirm the existence of an angular scale of transition to homogeneity, a result that strengths 
the CP.

\acknowledgments

FA, CPN, and AB acknowledge fellowships from CAPES, FAPERJ, and CNPq, respectively. 
EdC acknowledges the PROPG-CAPES/FAPEAM program. 
We would like to thank Joel Carvalho, Rodrigo S. Gon\c{c}alves, Jailson Alcaniz, and 
Roy Maartens for useful comments and productive feedback of our analyses. 
We acknowledge M. P. Haynes and the ALFALFA team for the use of the data.

\end{document}